\documentclass[aps,pra,twocolumn,amsmath,amssymb,showpacs,superscriptaddress,floatfix]{revtex4-2}
\usepackage{color}
\usepackage{textcomp} 
\usepackage{mathrsfs,amsmath}
\usepackage{graphicx}
\usepackage{dcolumn}
\usepackage[mathlines]{lineno}
\usepackage{hyperref}
\usepackage{bm}
\usepackage{epstopdf}
\usepackage{soul}
\usepackage{epsfig}
\usepackage{bbold}
\usepackage{braket}
\usepackage{physics}
\usepackage{gensymb}
\usepackage{amsmath,amssymb}
\usepackage{xcolor}
\usepackage{float}

\begin{document}
\title{Topological Control of Chirality and Spin with Structured Light} 

\author{Light Mkhumbuza}
\affiliation{School of Physics, University of the Witwatersrand, Private Bag 3, Wits 2050, South Africa}

\author{Pedro Ornelas}
\affiliation{School of Physics, University of the Witwatersrand, Private Bag 3, Wits 2050, South Africa}

\author{Angela Dudley}
\affiliation{School of Physics, University of the Witwatersrand, Private Bag 3, Wits 2050, South Africa}

\author{Isaac Nape}
\email{isaac.nape@wits.ac.za}
\affiliation{School of Physics, University of the Witwatersrand, Private Bag 3, Wits 2050, South Africa}

\author{Kayn A. Forbes}
\email{K.Forbes@uea.ac.uk}
\affiliation{School of Chemistry, University of East Anglia, Norwich Research Park, Norwich NR4 7TJ, United Kingdom}

\begin{abstract}
\end{abstract}

\begin{abstract}
Structured light beams with engineered topological properties offer a powerful means to control spin angular momentum (SAM) and optical chirality, key quantities shaped by spin-orbit interaction (SOI) in light. Such effects are commonly associated with non-paraxial focusing or light-matter interfaces. Here, we demonstrate that higher-order Poincaré modes carrying a tunable Pancharatnam topological charge~$\ell_{p}$ enable deterministic control of SOI entirely in free space and within the paraxial regime. We show that modulation of~$\ell_{p}$ drives a measurable radial separation of circular polarisation components – a free-space optical Hall effect arising from propagation-induced mechanisms alone. The effect originates from differential Gouy-phase evolution and radial divergence between the two circular components of an initially spin-balanced vector beam. This identifies~$\ell_{p}$ as a single, tunable parameter linking Pancharatnam topology to paraxial spin–orbit coupling, establishing a simple and material-independent route to generate and control optical chirality and SAM. This approach provides new opportunities for tunable optical manipulation, chiral sensing, and high-dimensional photonic information processing.
\end{abstract}

\maketitle
 
\section{Introduction}

Light’s spin and orbital angular momentum (SAM and OAM) arise from circular polarization and helical phase structure, respectively \cite{bliokh2015transverse}. Both are inherently chiral: SAM is defined by the helicity $\sigma = \pm1$, corresponding to right- and left-handed circular polarization, while OAM is quantified by the integer-valued topological charge $\ell = \pm1, \pm2, \ldots,\pm\infty$, with the sign indicating handedness and the magnitude the number of phase twists. Together, SAM and OAM form the total angular momentum of light, a central concept in modern photonics that underpin a wide range of applications \cite{forbes2024orbital, franke202230}, including classical and quantum communication \cite{shen2019optical, nape2023quantum}, light-matter interaction \cite{babiker2019atoms, quinteiro2022interplay, porfirev2023light}, and optical manipulation~\cite{padgett2011tweezers, yang2021optical}. Moreover, these degrees of freedom lie at the heart of structured light \cite{forbes2021structured, he2022towards}, i.e., custom-shaped light fields, that fuel a host of exotic phenomena ranging from the emergence of field textures that mimic particle-like topologies across quantum and classical domains, to exotic Berry-phase-driven effects that are enabled by spin-orbit interactions (SOI) \cite{bliokh2015spin, cohen2019geometric, shen2024optical}.

Traditionally, SOI in optical systems have been realized through structured light–matter interactions. These include anisotropic media such as patterned liquid crystals that exploit geometric phase~\cite{chen2020liquid}, metasurfaces that impose spatially varying polarization responses \cite{genevet2017recent, dorrah2022tunable, he2022towards, kuznetsov2024roadmap}, or spin-dependent surface waves in dielectric and plasmonic interfaces \cite{bliokh2015quantum, lodahl2017chiral, suarez2025chiral}. In tightly focused beams, SOI can emerge as spin-to-orbit conversion in scalar beams \cite{bliokh2010angular, bliokh2015spin} or give rise to phenomena such as the orbit-induced spin Hall effects of light \cite{fu2019spin, li2021spin, wu2024controllable}, where circular polarization components of vectorial fields can separate radially, occupying specific regions in the transverse plane of an optical field. However, to enhance the typically subwavelength and inherently weak SOI phenomena, one generally relies on non-paraxial conditions or tailored materials. This is primarily because these phenomena are significantly suppressed in the paraxial regime due to their minimal magnitude.

\begin{figure}[t]
\centering
\includegraphics[width= 0.9 \linewidth]{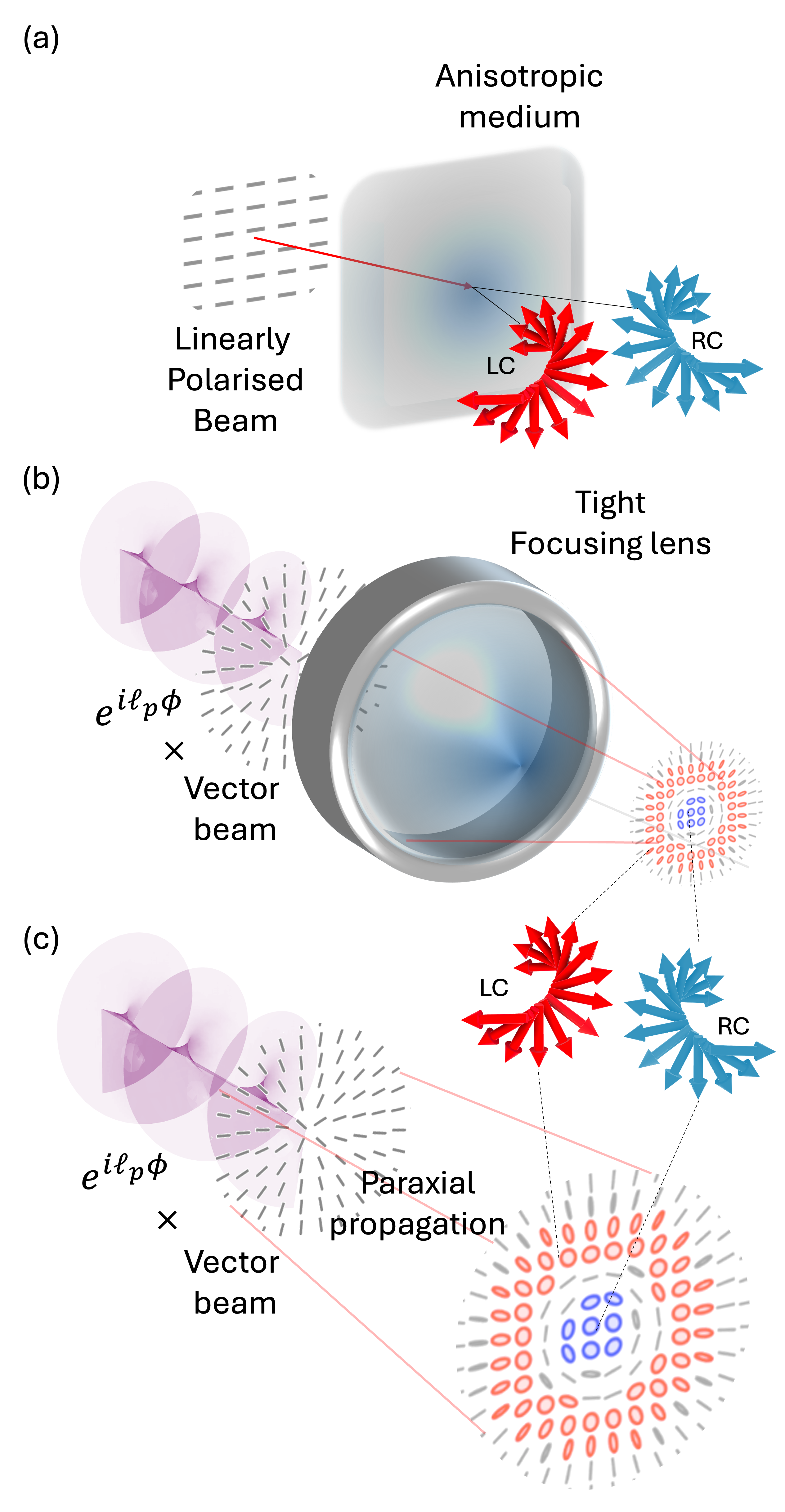}
\caption{\textbf{Concept of  spin-separation in vectorial fields.} (a) Spin-dependent separation resulting from a vectorial field interacting with an anisotropic medium, therefore resulting in a photonic spin-Hall effect. Lateral transverse shifts are observed depending on the spin components of the incident field. (b)
Spin separation induced by tight focusing of a radially polarized vectorial beam carrying a PT phase with a corresponding  topological charge, $\ell_p$,  and (c) reproduced by propagating the same field but through freespace without any interactions with matter. In each case, spin separation is observed in the evolved vector beam, showing regions dominated by right-handed and left-handed circular polarizations.}
\label{fig: Concept}
\end{figure}


Here, we demonstrate that the intrinsic topology of a vectorial beam can induce measurable spin–orbit interactions during free-space paraxial propagation. Specifically, we observe the emergence of local spin and optical chirality from an input field that is purely radially polarised and has $S_{3}=0$ everywhere at $z=0$. By encoding a Pancharatnam topological (PT) index $\ell_{p}$ onto the beam, which determines the global polarization-phase winding and the associated OAM \cite{niv2006manipulation}, the two circular components evolve with different Gouy-phase and divergence behaviour, producing controlled longitudinal spin without the need for tight focusing or material interfaces. This provides a simple, tunable means to generate and position chirality and spin densities in free space, with potential relevance to chiral sensing, optical manipulation, and high-dimensional information encoding.

While our results exhibit skyrmion-like coverage at later propagation planes, the present work does not seek to engineer optical skyrmions nor to reproduce previous paraxial skyrmion constructions \cite{gao2020paraxial, shen2025free}. In contrast to approaches that impose amplitude or OAM asymmetries at the source plane, such as rational-map designs \cite{cisowski2023building} or vector-field synthesis using metasurfaces \cite{hakobyan2024unitary, hakobyan2025q}, the input field studied here is a strictly symmetric, spin-balanced radial vector beam with $S_{3}=0$ everywhere at $z=0$. The only modification is the Pancharatnam topological index $\ell_{p}$, which does not affect the initial polarization distribution. The emergence of spin and chirality during propagation arises solely because $\ell_{p}$ forces the two circular components into different paraxial modal families with distinct Gouy-phase and divergence evolution. In this way, the Pancharatnam topological index is identified as a physically relevant control parameter for paraxial spin-orbit interaction in free space, establishing a previously unexplored connection between Pancharatnam topology and the generation of spin and chirality during propagation.

\section{RESULTS}

\textbf{Concept of spin-orbit interactions driven by beam topology.}
Before demonstrating how topological structuring enables SOI in paraxial light, we briefly outline their origins in light-matter interaction. Optical SOI manifest when initially linearly polarized beams develop spatially varying spin, typically via interaction with dielectric surfaces, anisotropic and inhomogeneous media, or tight focusing. These processes produce spin-dependent beam shifts such as the Goos-Hänchen and Imbert-Fedorov effects~\cite{bliokh2013goos, kim2023spin}, and radial spin separation driven by orbital structure~\cite{wu2024controllable}, all arising from spin-orbit coupling~\cite{bliokh2015spin}, as illustrated in Fig. \ref{fig: Concept}(a), (b), respectively.

Accordingly, these effects can also be viewed as being produced by polarization-dependent field gradients that separate spin components, akin to spin transport in electronic systems where electron spin-dependent flow and separation in the presence of an electric current can be observed \cite{cohen2019geometric}. As the spins separates spatially, two signatures emerge: (i) chiral spin textures with localized optical chirality (helicity), and (ii) spin currents characteristic of photonic spin-Hall effects. Here we show that by encoding topological structure into vector beams, these features can be realized in free-space within the paraxial regime – enabling spin separation and chiral spin flows without relying on strong focusing or complex media.\\

\textbf{Topology-driven spin-orbit interactions in paraxial light.}
To generate SOIs that result in spin generation and separation within laser beams, one can tightly focus (see Fig.~\ref{fig: Concept}~(b)) a scalar beam~\cite{yu2018orbit, kotlyar2020spin, forbes2021measures} or vector vortex beam~\cite{han2018catalystlike, li2018orbit, shi2018structured, forbes2024spin, meng2019angular, li2021spin, geng2021orbit, fang2021photoelectronic, zhang2022ultrafast, man2022polarization, liu2024manipulation, wu2024controllable} to produce so-called orbit-induced local spin (OILS). However, as we will demonstrate, these effects can also be realized in the paraxial regime, without the need for tight focusing. The approach begins with an optical field that exhibits a spatially varying linear polarization combined with a global azimuthal phase profile, as illustrated (c). The corresponding  electric field has the initial profile
\begin{equation}
\mathbf{U}_{\text{in}} (\textbf{r}) \propto e^{i\ell_p \phi} \times \underbrace{ f_{\text{in}}(\mathbf{r}) \left(  \text{e}^{i\Delta\ell\phi} \hat{\sigma}_{+} + \text{e}^{-i\Delta\ell\phi} \hat{\sigma}_{-}\right)}_{\text{radial field} },
\label{eq:HOPs}
\end{equation}
 in polar coordinates, $\textbf{r} = (r, \phi)$  with $\sigma_\pm $ representing the right and left CP, each marked with azimuthal phase profiles, $\exp(\pm i\Delta\ell \phi)$  with $\Delta \ell \in \mathbb{Z}$.  At this point, it is crucial to notice that the polarization components each have the same radial amplitudes ($f_{\text{in}}(\cdot)$). The global phase, which has been factorized from the state,  $\exp(i\ell_p \phi)$  characterized by the PT  index $\ell_p\in \mathbb{Z}$, encodes the elusive topological features that we are interested in as this controls the spin-orbit interactions in the field. Given the above electric field, the topological phase can be computed from (See Supplementary Material) 
\begin{equation}
    \phi_p=\arg(\langle\mathbf{U}_{\text{in}}(0)|\mathbf{U}_{\text{in}}(\phi)\rangle)=\frac{\ell_p}{2}\phi, 
\end{equation}

\noindent which is directly linked to the PT winding number that is also associated with the total OAM of the VVB (see Supplementary Material). Setting $\ell_p =0$ produces a typical radially polarized field that does not carry this phase (nor total OAM) and remains unchanged, i.e., maintaining the same polarisation profile on propagation. However, for $|\ell_p|>1$ the fields carries nonzero OAM while the PT phase causes the polarisation field to change across the beam in the transverse plane on propagation, so that the new field maps as
\begin{equation}
\mathbf{U}_{\text{in}} \rightarrow \mathbf{U} = f_{\ell_{\text{A}}}(\mathbf{r})  \text{e}^{i\ell_\text{A}\phi} \hat{\sigma}_{+} +  f_{\ell_{\text{B}}}(\mathbf{r})  \text{e}^{i\ell_\text{B}\phi} \hat{\sigma}_{-},
\label{eq:HOPs2}
\end{equation}
\noindent where  $f_{\ell_{\text{A(B)}}} (\mathbf{r})$ represent the new field amplitude distributions for each spin component, that now depend on the PT charge,  $\ell_\text{A(B)} = \ell_{p}\pm \Delta \ell$, respectively.  As a consequence, one observes that the spin components separate radially in the field. For example, in the output profile of the concept illustration in Fig. \ref{fig: Concept} (c), we see that close to the core ($r \approx 0$), the field is dominated by right circular spins and subsequently dominated by left circular spins as one scans the field radially outward - a key signature of local spin generation  - whereas initially the field had no local spin components. This can be measured by quantifying the local optical chirality density and longitudinal spin densities, which are both proportional to the third stokes parameter (See Supplementary Material for the derivation), $S_3$ \cite{forbes2024spin}, i.e,
\begin{align}
S_3=|f_{\ell_\text{A}}(r,z)|^2-|f_{\ell_\text{B}}(r,z)|^2,
\label{eq:chirality}
\end{align}
that can be extracted from the transverse plane for various longitudinal coordinates, $z$. The third Stokes parameter can be measured experimentally via the difference in intensities between the space-dependent amplitudes marking each spin state.  The reason for extracting the optical chirality $C$ and spin density $\mathbf{s}$ from the stokes parameter ($S_3$) follows from the fact that in the paraxial regime, these quantities take familiar forms, with the spin being purely longitudinal; $s_z \propto C\propto \sigma I$,  where $\sigma$ is the helicity (for circular polarization $\sigma = \pm 1$, for elliptical $|\sigma| < 1$, and for linear or unpolarized light $\sigma = 0$), and $I$ is the beam intensity.
We show that radial spin separation originates from amplitude symmetry breaking due to topological encoding, and elucidate its propagation dependence via $S_3$ and the associated polarization profiles.

\textbf{ Inducing SOI in paraxial fields through topology-driven amplitude change in propagation.}  
\begin{figure*}[t!]
    \centering
\includegraphics[width=\linewidth]{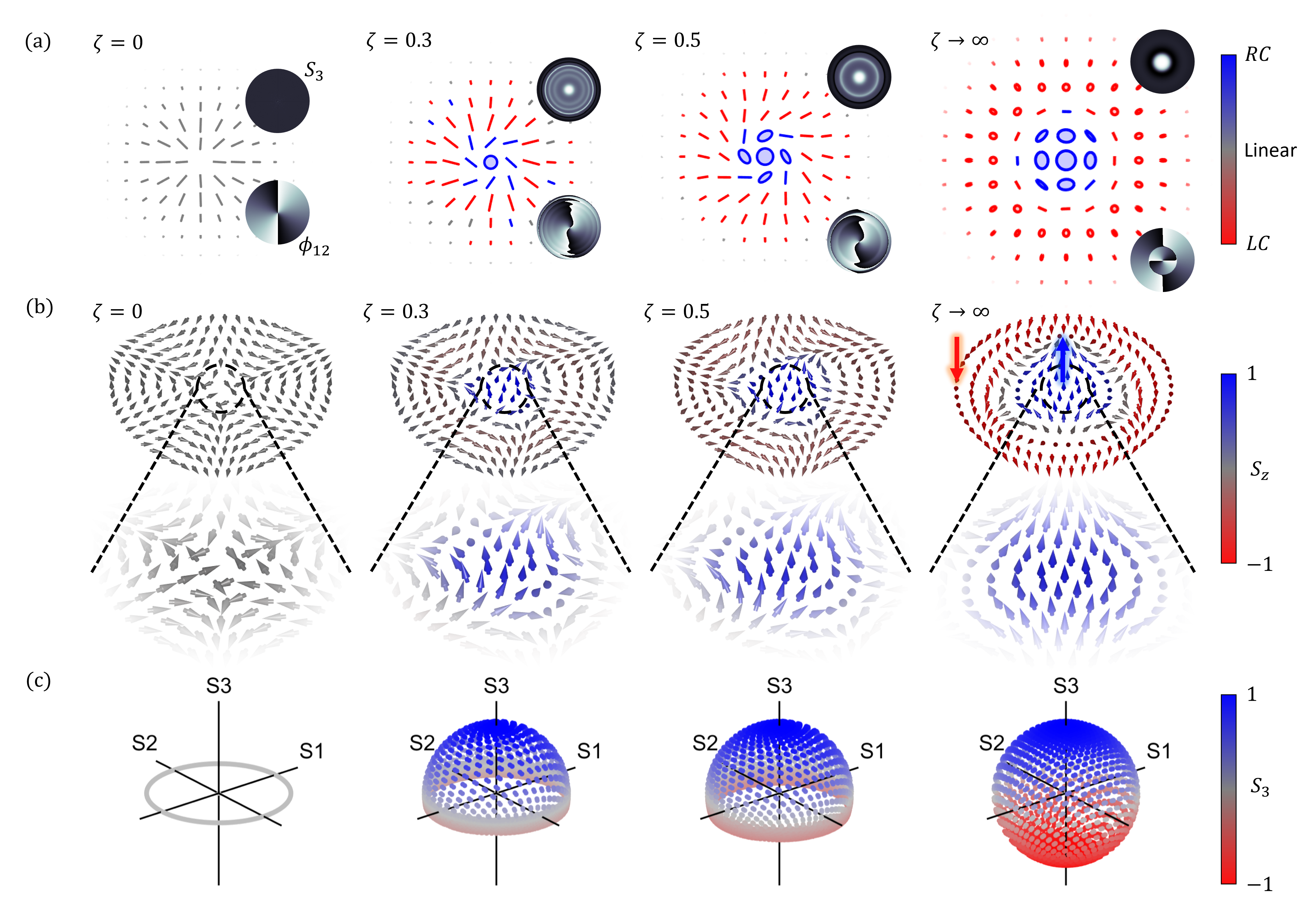}
    \caption{\textbf{Revealing the origin of orbit-dependent spin dynamics in paraxial light}. (a) A horizontally polarized LG mode (with $\ell_p = 1$ and $p = 0$) incident on a $q$-plate, produces a vector mode with a radially polarized field pattern (in the near field) but carrying a net OAM charge of $\ell_p$.  Initially, the field contains zero spin density as it is populated by linear polarization states. On propagation, the beam shows  a varying chirality and spin density ($S_3$), shown via the polarization ellipses,  at various propagation planes.   These propagation planes are marked by the ratio $\zeta = z/z_R$, with $\zeta=0$ corresponding to the image plane. The spin density ($S_3$, top panel) and the relative phase ($\phi_{12}$, bottom inset) are shown at each propagation plane. (b) Spin textured fields represented by spin unit vectors for the various corresponding propagation planes, with selected zoomed-in regions.  Initially, the spin vectors point in the transverse plane, then gradually accumulate upward right-circular (RC) and downward left-circular (LC) spin components at the center and away from the origin, respectively. In the far field, there is a clear boundary that separates the LC and RC spin components, indicative of the Hall effect - orbit dependent spin separation. (c)  The population of polarization states is shown at each plane on the Poincaré sphere.   Initially, only the equator is covered at the waist plane, since the field contains only linear polarization states. Eventually, full coverage is shown, illustrating that the field evolves into a full Poincaré beam.}
    \label{fig:propagation}
\end{figure*}
\begin{figure*}[t]
    \centering
\includegraphics[width=\linewidth]{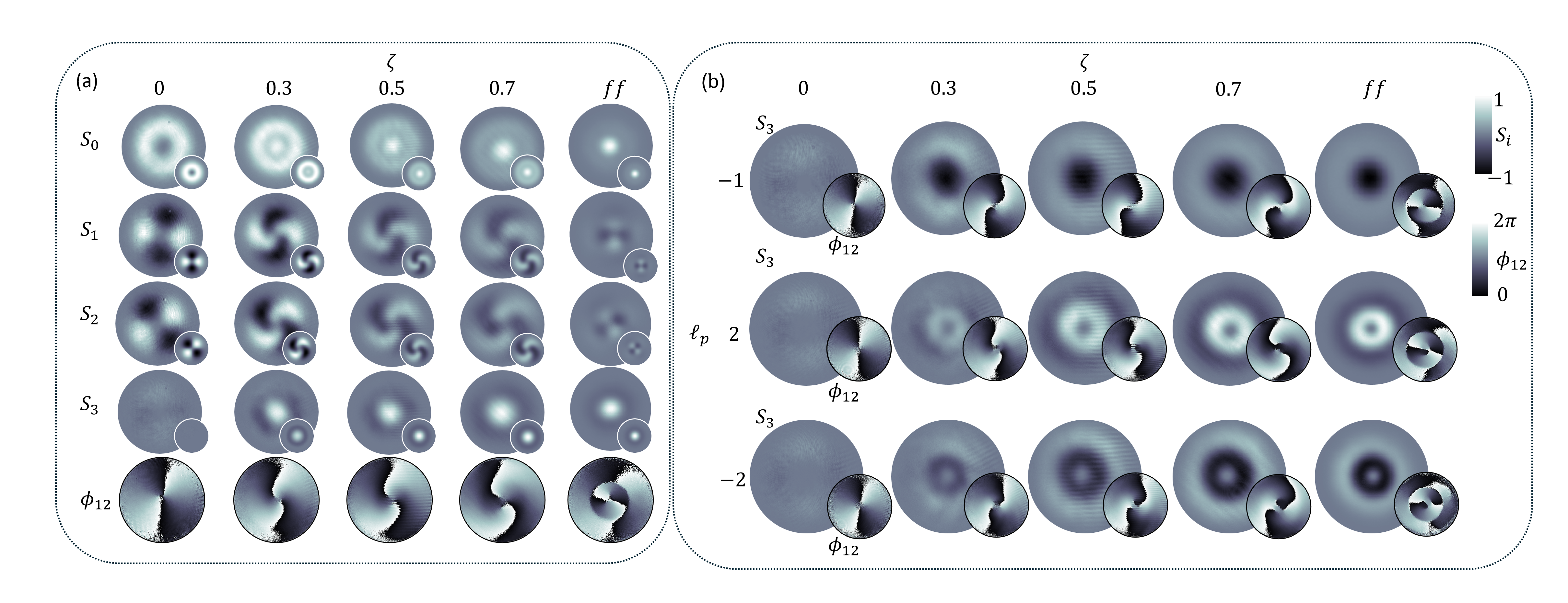}
    \caption{ \textbf{Experimental stokes parameter analysis.} (a) Measured Stokes parameters, $S_j$, for $\ell_p=1$ at various propagation planes, labeled here as the ratio $\zeta = z/z_R$ with $\zeta=0$ corresponding to the image plane. The last row shows the relative phases, $\phi_{12}$, showing a winding number  $2 \Delta \ell \approx 2$  for all propagation planes. (b) The measured $z$ component of the Stokes parameter $S_3(\mathbf{r})$, characterizing the spin density of the field at different planes, $\zeta$ and the corresponding relative phase ($\phi_{12}$,  see inset). This is shown for $\ell_p = -1, 2$ and $-2$.}
    \label{fig:Exp_figure2}
\end{figure*}

\begin{figure*}[t]
    \centering
    \includegraphics[width=0.9\linewidth]{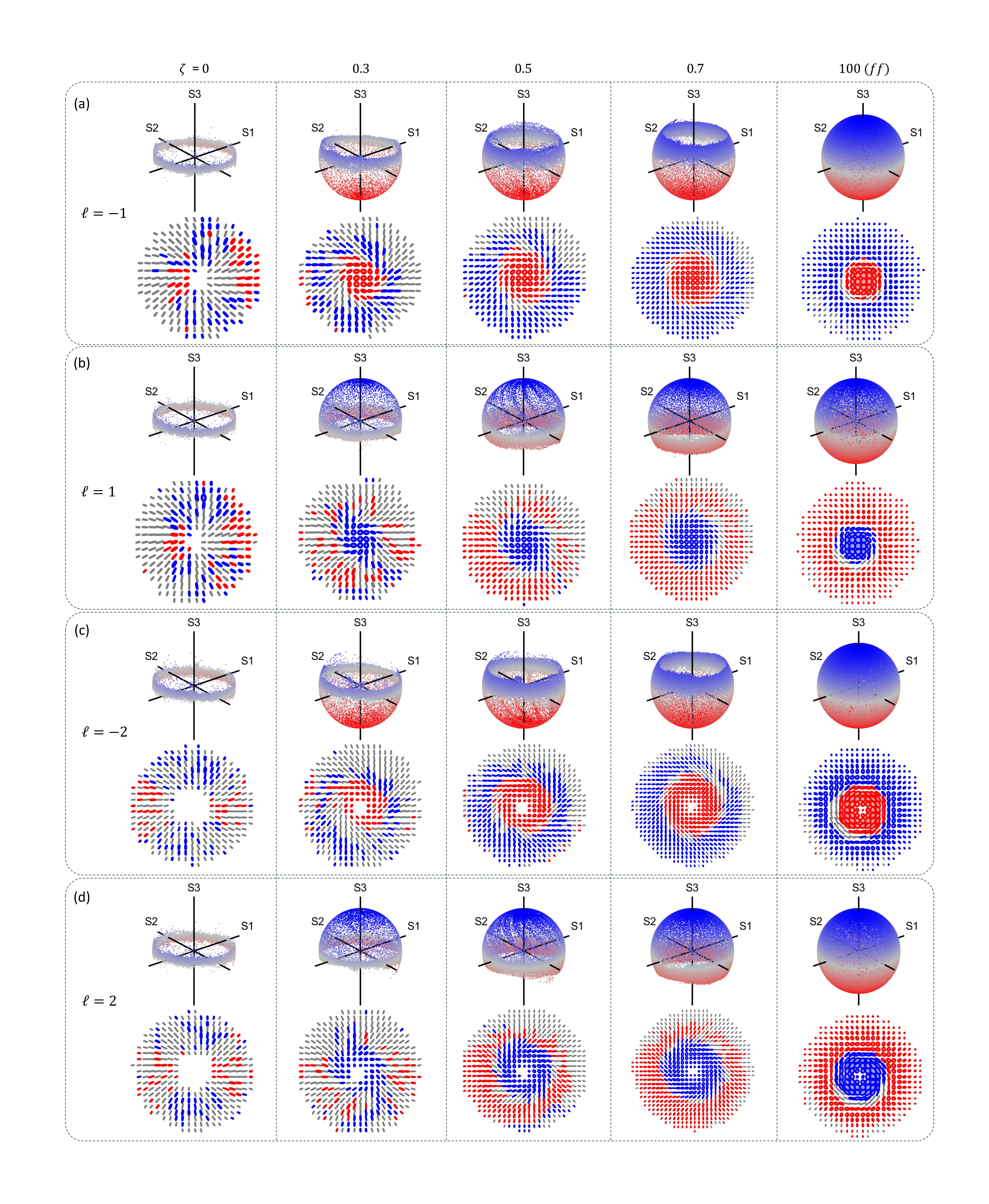}
    \caption{\textbf{Observing orbit induced spin induction upon propagation}. Reconstructed Poincaré sphere coverage and polarization ellipses for Pancharatnam topological indexes (a) $\ell_p = -1$  and (b) $\ell_p = 1$  at various propagation planes ($\zeta = z/z_R$). The same plots are shown for topological phases (c) $\ell_p=-2$ and (d) $\ell_p=2$, illustrating the emergence of spin from a vectorial field that initially has zero spin density in the paraxial regime.}
    \label{fig:Exp_figure3.1}
\end{figure*}
Now we uncover the mechanism that enables one to observe SOI in paraxial beams. Firstly, notice that the vector beams described by Eq. \ref{eq:HOPs2} are hybrid-order Poincaré beams (HyOPs)  and are not eigenmodes of free-space propagation \cite{sanchez2019gouy} when $|\ell_p|>0$.  As they propagate, the local states of polarization evolve along $z$ due to differential Gouy phase shifts and radial amplitude variations between the constituent modes that mark the right and left CP states. This leads to a spin-dependent splitting into concentric rings, each carrying opposite circular polarization and different OAM – a signature of propagation-driven optical Hall effect. While hybrid-order vector beams are not eigenmodes of free-space propagation, this alone does not fix the form of the resulting spin density; here, its emergence is deterministically governed by the Pancharatnam topological index.

To demonstrate this in the paraxial regime, we prepared a scalar horizontally polarized Laguerre-Gaussian (LG) mode, as shown on the left of the concept image in Fig. \ref{fig: Concept} (c), having a characteristic field profile that can be expressed as 
\begin{align}
\text{LG}_{\ell_p}(\mathbf{r}) \hat{\mathbf{x}} &= f_{\ell_p}(r,z) \text{e}^{i\ell_p\phi} \hat{\mathbf{x}},
\end{align}
 where $(\hat{\mathbf{x}} = \hat{\mathbf{\sigma}}_{+} +  \hat{\mathbf{\sigma}}_{-})/\sqrt{2}$ and $\mathbf{r} = (r, \phi, z)$ are the polar coordinates. The LG modes used here have a characteristic radial profile,  $  f_{\ell_p}(r,z) \propto (\sqrt{2}r / w)^{|\ell_p|} \exp(i(\psi_\text{G}+kz)-r^2/w^2)$  where $w[z]$ is the waist size of the embedded Gaussian component of the field ($\text{e}^{-r^2/w^2}$); a Gouy phase $\psi_\text{G}=(|\ell_p|+1)\arctan(z/z_\text{R})$; and an azimuthal phase,  $\exp(i\ell_p \phi)$. Note how the topological charge determines the change in the Gouy phase term, as well as the divergence through the radial term.  The resulting field will be modulated with a spin-orbit coupling (SOC) device, e.g., a $q$-plate, at the $z=0$ plane, yielding the mapping
\begin{align}
\text{LG}_{\ell_p} \hat{\mathbf{x}} & \xrightarrow{q \text{-plate}}  \mathbf{U}_{\text{in}}(\mathbf{r}, z=0), \nonumber \\ 
 &= f_{\ell_p}(r) \left( \text{e}^{i \ell_\text{A} \phi} \hat{\sigma}_{+} + \text{e}^{i \ell_\text{B} \phi} \hat{\sigma}_{-} \right), \label{eq:LGbeam}
\end{align}
therefore producing the HyOP in Eq. (\ref{eq:HOPs})  but at the $z=0$ plane,  with  $\ell_\text{A} = \ell_p + \Delta \ell$ and $\ell_\text{B} = \ell_p - \Delta \ell$ while $| \Delta \ell|$ is the magnitude of the topological charge transferred by the $q$-plate for each spin component, i.e. $\pm \Delta \ell$ for $\sigma_{\pm}$, respectively.  In this work, $\Delta \ell = 1$ due to the $q$-plate. While the $q$-plate is employed to generate the radially polarized mode, we emphasize that it is not fundamental to the SOI effect under investigation. It merely serves as a convenient means to prepare the desired beam, and alternative methods could be equally suitable, e.g., using dynamic phase control with spatial light modulators and interferometers.

At this stage, the field in Eq. (\ref{eq:LGbeam}) is a typical vectorial field that is linearly polarized and carries the PT phase, which at $z=0$, has no impact of the field. 
One would expect the field to evolve like a cylindrical vortex beam that maintains its polarization profile upon propagation,  however, the global phase, $\exp(i\ell_p \phi)$, carrying the PT phase, causes an asymmetry in the amplitudes. Theoretical simulations of the resulting polarization ellipse profiles are depicted in Fig.~\ref{fig:propagation}~(a), illustrating the change in chirality (accumulation of CP components) as the resulting mode propagates.  These are obtained using the localized Stokes vector ($\mathbf{S} \left( \mathbf{r} \right) = \langle S_1(\boldsymbol{r}), S_2(\boldsymbol{r}), S_3(\boldsymbol{r}) \rangle$) components (See Methods for Stokes parameter reconstruction procedure). Here, the propagation planes, $\zeta = z/z_\text{R}$ , are defined relative to the Rayleigh range ($z_\text{R}$).
The spin densities ($S_3$) are shown as the top insets of Fig. \ref{fig:propagation} (a), confirming the increase in spin density while the relative phase ($\phi_{12} = \arg \left( S_1(\mathbf{r})+i S_2(\mathbf{r}) \right) = 2 \Delta \ell \phi $, from which the polarization order ($\eta = \Delta \ell/2 $) can be determined,  is preserved, shown in the bottom inserts of Fig.\ref{fig:propagation} (a).
The spin textures, illustrating the directional spin unit vectors, satisfying $\sqrt{ S_1^2 (\mathbf{r}) + S_2^2 (\mathbf{r}) + S_3^2 (\mathbf{r})} = 1$, are shown in Fig. \ref{fig:propagation} (b). Here we see that initially at the $\zeta = 0$ (equivalently $z=0$) plane, the third Stokes parameter evaluates to zero at all positions of $\mathbf{r}$ as shown in Fig. \ref{fig:propagation} (a) which is reflected in  Fig. \ref{fig:propagation} (b) where the spin vectors are all in plane. This is because the spatial amplitudes are the same ($|f_{\ell_\text{A}}|^2 = |f_{\ell_\text{B}}|^2 = |f_{\ell_p}|^2$ )  for the spin components. Therefore, the field is not chiral at this specific plane ($z =0$). However, this is not the case for $\zeta>0$.\\

The position-dependent amplitudes have been derived analytically for cases where an LG mode is incident on a phase element that imparts a phase of $\text{e}^{\pm i\Delta \ell \phi}$ similar to the phases imparted by the  $q$-plate. It has been shown that the LG mode profile evolves into an elegant Laguerre-Gaussian (eLG) \cite{cocotos2025laguerre} mode or, equivalently, into a Hypergeometric Gaussian (HGG) mode \cite{karimi2007hypergeometric}. In this article, we will use eLGs to represent the modes as they evolve, i.e.~eLG$^{\ell_{\text{A(B)}}}_{p_{\text{A(B)}}}$,  with a topological charge of $\ell_{\text{A(B)}} = \ell_p \pm \Delta \ell$  and a radial index $ p_{\text{A(B)}} = \frac{1}{2} \left( \abs{\ell_{p}} -\abs{\ell_\text{{A(B)}}} \right)$~\cite{cocotos2025laguerre}. These quantum numbers explicitly depend on the Pancharatnam index, $\ell_p$. Therefore, the amplitude changes in the polarization components of the fields satisfying $f_{\ell_\text{A(B)}}(r,z)\equiv |\text{eLG}^{\ell_{\text{A(B)}}}_{p_{\text{A(B)}}}(r,z)|$, are responsible for the observed polarization profile changes that in turn produce areas of chirality and longitudinal spin (i.e.: $S_3 \neq 0$) in the propagated fields.  The angular spectrum (equivalently the farfield amplitude profile) of these modes has the form (see Supplementary Material for complete expression),  \begin{align}
\text{eLG}^{\ell_{\text{A(B)}}}_{p_{\text{A(B)}}}(\rho, z\rightarrow\infty) & \propto i^{-\ell_\text{A(B)}}  \rho^{|\ell_\text{A(B)}|} L^{|\ell_\text{A(B)}|}_{p_\text{A(B)}}[\rho^2] \nonumber \\ &\quad\times \exp(i \ell_\text{A(B)} \phi),
\end{align} where $\rho$ is the normalized radial (wavenumber) coordinate showing the explicit dependence on the radial term $\rho^{|\ell_\text{A(B)}|} = \rho^{|\ell_{p} \pm \Delta \ell|}$ which controls the radial distribution of the modes (see Supplementary Material for more details about radial amplitude separation). Therefore, this radial factor  is the source of the asymmetry in the spin components that can be observed in the  transverse plane of the fields.
In Fig. \ref{fig:propagation} (c) we map the fields onto the Poincaré sphere and show the population of polarization states across a field. Initially, the field at $z=0$, maps onto a ring on the equator of the sphere because the field has no longitudinal spin components ($S_3 = 0$). However, as the field propagates, gradual coverage over the sphere is achieved, indicating the emergence of a longitudinal spin component ($S_3 > 0$) in the fields. Furthermore, because states near the origin have a higher intensity, only one half of the Poincaré sphere is covered first and other states are gradually populated as the beam propagates. Eventually, the farfield ($z\rightarrow \infty$) is fully occupied by all possible spin states. This is observed taking into account that the intensity of the fields in the radial direction decreases, and nearly about $8 \%$ of the peak intensity is detected with a typical off-the-shelf CCD camera (Thorlab's CS505MUP1) under the influence of stray light and shot noise in the detector (CCD). In the Supplementary Material, we show the case where the entire field intensity can be resolved completely, and find that full coverage is observed immediately upon propagation; however, we use the former to mirror experimental conditions.\\

\textbf{Experimental Validation}
Next, we performed an experiment to validate the above  (See experimental details in Methods). Horizontally polarized LG modes were prepared sequentially (using a spatial light modulator) with varying topological charges, $\ell_p \in \{1,-1, 2, -2\}$,  with each mode imaged onto a $q$-plate that transfers a net charge of $\pm\Delta\ell = \pm 1$ for each circular polarization component $\sigma_{\pm}$, respectively.   This produces our HyOP mode, carrying a Pancharatnam topological charge $\ell_p$.

The Stokes parameters are shown in Fig. {\ref{fig:Exp_figure2}} (a) at various propagation planes ($\zeta = \{0, 0.3, 0.5, 0.7, \text{farfield }(ff)\})$  first for $\ell_p=1$, with the last row showing the relative phase $\phi_{12} $ .  Although the $S_1$ and $S_2$  Stokes parameters have the same number of lobes  as shown in the two middle columns (second and third rows) of Fig. \ref{fig:Exp_figure2} (a), the only change observed is in the spiralling nature of the pattern upon propagation due to curvature and relative Gouy phase between the evolving field's components. The twist of the spiral phase observed in the relative phase ($\phi_{12}$, last row of Fig. 
\ref{fig:Exp_figure2} (a)) is seen to depend on the sign of the Pancharatnam topological charge during propagation. However, its handedness (direction of increasing phase in the azimuth direction) is preserved. On the other hand, the total beam intensity ($S_0$) and spin densities ($S_3$)  are seen to change upon propagation as predicted, due to the amplitude change, with the left-handed CP component occupying the centre. In contrast, the right-handed CP components dominate regions further away from the centre. 

Next, in Fig. \ref{fig:Exp_figure2} (b) we varied the input Pancharatnam topological charge $(\ell_p \in \{-1, 2, -2\})$ and show the spin densities ($S_3$) and the relative phases $\phi_{12}$.  All the cases also exhibit identical relative phase profiles, although the spin densities ($S_3$) differ, resulting in different amplitude responses. In contrast to $\ell_p =1$, $\ell_p = -1$, it  produces a spin density profile, $S_3$, that is inverted, showing that the chirality in the field has switched; now the right-handed CP amplitudes dominate in the center, whereas the left-handed CP components dominate at a distance away from the centre.  For $\ell_p=\pm 2$ we also see the same trend where the dominant polarization switches depending on the input topological charge. Further, the magnitude of the input charge ($\ell_p$) is seen to control the radial profile of the spin density ($S_3$) owing to the radial modes that emerge in the field upon propagation, since the field components become eLG modes that have regions of nonoverlapping intensities. Moreover, for $\ell_p = \pm 2$, a vortex is observed since both circular polarization components have non-zero OAM after passing through the $q$-plate, which can also be inferred from the $S_3$ components.  We emphasize that the same $q$-plate was used to obtain these results, indicating that the local chirality is primarily attributed to the presence of Pancharatnam topological charge in the fields.

The sphere coverage and polarization (ellipse) profiles at different propagation planes are depicted in Fig. \ref{fig:Exp_figure3.1}. The polarization ellipses are represented in terms of handedness, with red indicating left-handed CP and blue representing right-handed CP.

Results were obtained for positive topological charge $\ell_p > 0$ (middle and bottom rows of Fig. \ref{fig:Exp_figure3.1}), resulting in right-handed CP at the origin of the field. Conversely, for $\ell_p < 0$ (top row of Fig. \ref{fig:Exp_figure3.1}), left-handed CP was observed at the center/origin of the field. We observe that the fields only contain linearly polarized states at the plane of the $q$-plate, as expected, since only linear polarizations are observed, indicated by the population of the fields on the equator of the Poincaré sphere. This can be inferred from the radially oriented linear polarization states, indicating that $S_3 \approx 0$. The presence of a few observed ellipses at this plane can be attributed to experimental errors induced by the waveplates. This can be improved through careful calibration of the waveplates. As we observe propagation, circular polarizations begin to form, indicating an increase in the spin density of the field. For $\ell_p=1 (-1)$, the accumulation of elliptical polarization states is also seen through the occupation of states beyond the equator. In fact, the polarization states initially dominate  the top half of the sphere, since the LC (RC) components have a larger amplitude contribution and eventually cover the whole sphere once the beam approaches the farfield . Similarly, this is observed for $\ell_p=\pm 2$, however, the structure has a characteristic vortex at the centre.
\begin{figure*}[t!]
    \centering
    \includegraphics[width=0.9\linewidth]{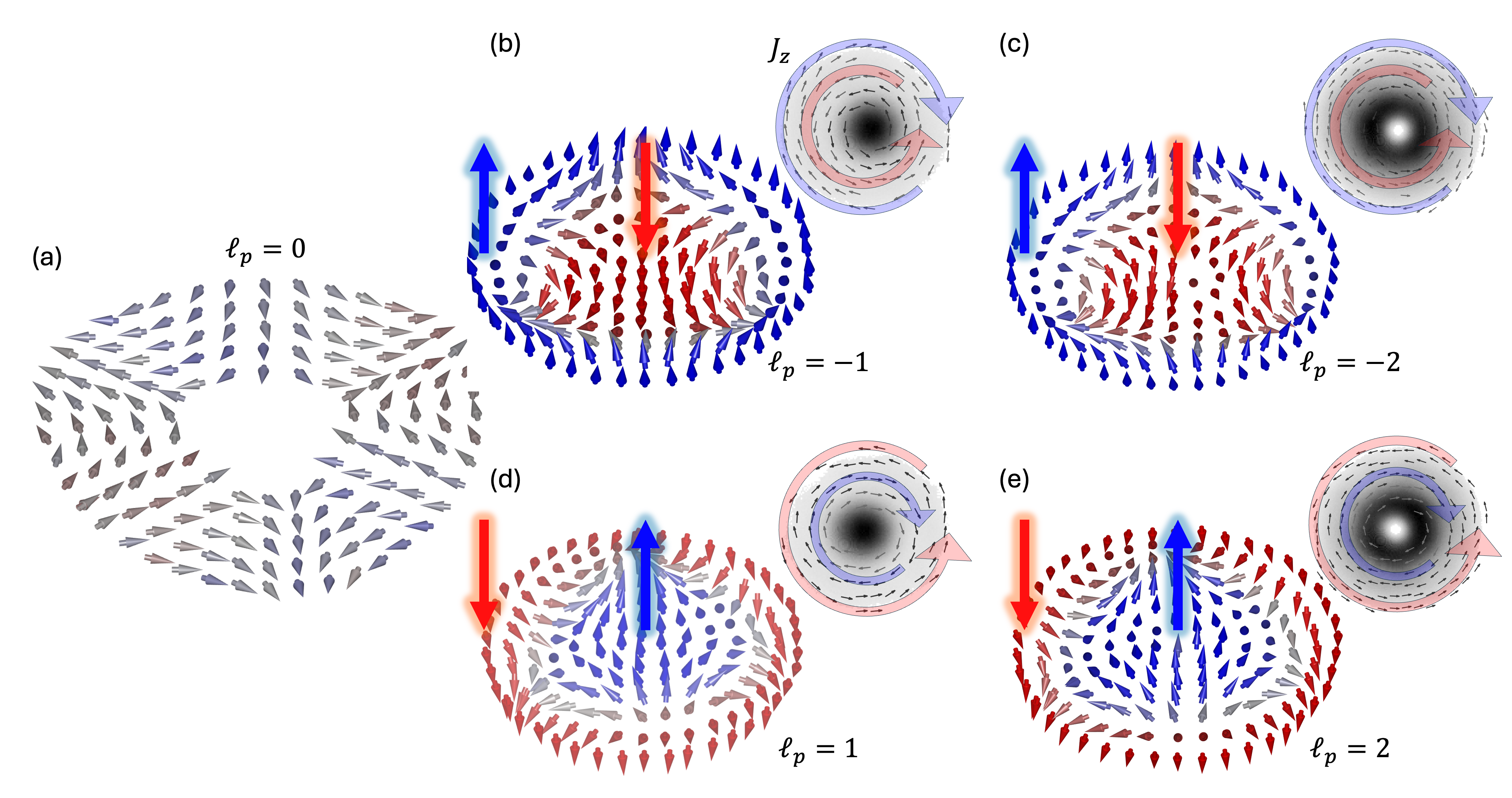}
    \caption{\textbf{Emergent orbit-induced Hall effect}. Experimental spin textures for the farfield modes given the Pancharatnam topological charge, $\ell_p =$ of (a) $0$, (b) $-1$, (c) $-2$, (d) 1 and (e) 2,  illustrating an orbit-induced 
    Hall effect. The transverse spin separation is visualized by the spin-current insets, which reveal an azimuthal flow that reverses direction as it approaches the field’s origin. This behaviour shows that the polarization handedness is well-defined near the center and changes beyond a certain radial distance, indicative of an optical spin-Hall effect. The insets show the spin currents ($\bar{J}$) which reveal the azimuthal spin currents due to the spin-Hall effect.}
    \label{fig:Exp_figureSpin}
\end{figure*}

\textbf{Topology-driven optical 
Hall effect}.  
Lastly, in Fig. \ref{fig:Exp_figureSpin}, we show the measured farfield spin textures for each of the HyOP modes having  different Pancharatnam topological charges.  In these illustrations, it can be seen that for $\ell_p=0$ all the spin vectors lie in the transverse plane, showing that there is no spin separation. However, once  $\ell_p>0$, the separation of RC and LC components is observed, confirming the topologically driven nature of the spin separation in the optical field, which only depends on $\ell_p$. Therefore, the initial Pancharatnam topological charge, which characterizes the net OAM of the  modes, induces transverse spin currents analogous to an electric field that induces similar currents in a magnetic system. The spin currents can be measured from $\bar{J} = \langle  \pdv{S_3}{y} ,   -\pdv{S_3}{x}  \rangle$. We show this for our experimetal results in the insert of Fig. \ref{fig:Exp_figureSpin}, demonstrating an azimuthal spin current in each case. This is because of the radial spin gradient seen in the $S_3$ components of the field with an abrupt switch in helicity observed in each case. In fact, more generally, this Hall effect can be interpreted as a spin-orbit Hall effect arising from the spatial separation of spin and orbital angular momentum components in vector vortex beams during propagation. For example, the inner and outer rings of Fig. \ref{fig:Exp_figure3.1} (c) in the $ff$ possess an azimuthal phase of $\text{exp}(i\phi)$ and  $\text{exp}(i3\phi)$, respectively. Although we have not directly measured the OAM content of each polarization component, the fact that their regions do not overlap suggests that each mode may contribute an OAM density confined to its respective nonoverlapping area.
Finally, it is important to emphasise that spin and optical chirality in these beams are local properties. Due to the divergence of the full three-dimensional integrals, the (integral) total values of the spin and optical chirality for beams are instead expressed in terms of their linear densities, i.e., the corresponding quantities per unit length along the $z$-axis: $\propto\int{S_3}d\mathbf{r}_\perp=0$. Clearly upon free-space propagation of the radially polarized field the integral values are conserved.

\section{Discussion}

Spin-dependent intensity redistribution during propagation of asymmetric OAM superpositions is known in structured-light research \cite{ling2014realization, zhang2015unveiling, li2015experimental, li2015modulation}. However, the physical situation considered here is different. The input beam carries no initial spin, no amplitude imbalance between $\sigma^{+}$ and $\sigma^{-}$, and no engineered OAM asymmetry: both circular components have identical radial amplitudes at $z=0$. The observed spin and chirality therefore do not come from pre-imposed modal asymmetry but arise exclusively from the Pancharatnam topological index $\ell_{p}$, which determines the distinct paraxial modal families into which the two spin components evolve. This isolates a propagation-induced SOI channel that has not been identified in previous demonstrations, where the spin splitting is already encoded at the generation plane. Importantly, this behaviour is not a generic consequence of paraxial propagation but is deterministically governed by the Pancharatnam topological index $\ell_p$. Our results show that a topologically charged but spin-free vector beam can self-generate measurable spin and chirality in free space, with $\ell_{p}$ acting as a deterministic and tuneable control parameter.

The emergence of non-zero spin and optical chirality densities (quantified by $S_3$)) from an initially spin- and chirality-free beam exemplifies orbit-induced local spin – OILS. While OILS has traditionally been linked to scalar \cite{yu2018orbit, kotlyar2020spin, forbes2021measures} and vector vortex beams \cite{han2018catalystlike, li2018orbit, shi2018structured, forbes2024spin, meng2019angular, li2021spin, geng2021orbit, fang2021photoelectronic, zhang2022ultrafast, man2022polarization, liu2024manipulation, wu2024controllable} in the non-paraxial regime, our findings show that, for vector vortex beams, OILS can also occur in the paraxial regime via a topologically driven optical Hall effect of light \cite{bliokh2015spin}, which was previously believed to require tight focusing. For instance, in a recent experiment of Wu \textit{et al.} \cite{wu2024controllable}, a linearly polarized Gaussian beam is first modulated using a spatial light modulator to produce a linearly polarized vortex beam. This was then passed through an $S$-plate to generate radially polarized light, which is subsequently tightly focused. However, the spatial separation of $S_3$ components observed in their experiment can be attributed to the focusing-independent, topologically driven mechanism we describe. While tight focusing may be necessary to produce measurable mechanical effects, such as particle manipulation, it is not the fundamental origin of OILS in vector vortex beams. Instead, we demonstrate that the underlying effect is paraxial and arises from the optical Hall effect of structured light. The intrinsic non-paraxial form of OILS arises from higher-order transverse electric field components whose magnitude scales as $(1/kw_0)^2$, making the associated spin generation extremely weak unless the beam is very tightly focused~\cite{forbes2026vortex}. In contrast, the Pancharatnam-driven paraxial OILS mechanism reported here originates entirely from the zeroth-order paraxial electric field and therefore produces a significantly stronger and more readily observable spin-orbit coupling effect (see Supplementary Information).

We have shown that the Pancharatnam topological charge $ \ell_p $ is responsible for inducing optical chirality and spin angular momentum in the paraxial regime. To demonstrate this, we employed a radially polarized beam with non-zero net OAM and a polarization order of $ \eta = 1 $. Although the polarization order remains invariant during propagation, the amplitude distribution of this mode evolves, giving rise to spin and chirality, as visualized through the Stokes parameter $ S_3 $. By varying the Pancharatnam topological charge $ \ell_p $, via an optical Hall effect of paraxial light, we demonstrate control over the resulting optical spin and chirality densities. We emphasize that the SOC device sets the OAM difference $\ell_\text{A} - \ell_\text{B} = 2 \Delta\ell$ (and equivalently the mode order, $\eta$) between the two spin components which drove the separation between the RCP and LCP components upon propagation. Choosing an SOC device with a larger $\Delta\ell$ (and $\ell_p \neq 0$) leads to a more pronounced Hall effect. This is due to the fact that the ring size of these modes behave roughly according to $r^{|\ell_p \pm  \Delta \ell|}$, which means that the greater the difference between the magnitude of the OAM values for each polarization component, the larger their radial separation in the far-field. While we employed an SOC device for its convenience in generating such beams, the underlying SOI effect is not unique to this method and can also be realized using alternative beam-shaping approaches such as spatial light modulators or interferometric schemes. Further, we emphasise that our beams, away from the source plane, exhibit transverse profiles similar to those of optical skyrmions \cite{zhang2025topological, wang2024topological, chen2025more}. While for skyrmions, the spin density is engineered at the source via amplitude shaping, in our case it is generated by the embedded  PT charge and phase and emerges upon propagation 


This work reveals a previously underexplored mechanism for spin–orbit interactions that generate local spin and chirality in paraxial structured light, governed by the Pancharatnam topological charge~$\ell_{p}$. Beyond providing new physical insight into free-space SOI, the mechanism offers direct utility for systems where tunable spin and chirality through~$\ell_{p}$ alone, without tight focusing, enable new practical applications. Such control allows dynamically adjustable chiral sensing, enhanced enantioselective light-matter coupling, and polarisation-tailored optical trapping and manipulation that benefit from tunable spin- and chirality-density hotspots and from free-space control of spatially separated spin modes. The ability to encode and tune coupled SAM-OAM states through a single topological parameter~$\ell_{p}$ provides a compact, material-independent route to generate structured spin-orbit photonic states for both quantum and classical information processing. Furthermore, the ability to predict when a given vector vortex beam will self-convert into a specific spin-orbit configuration offers a valuable tool for the design of SOI converters and structured-light sources. We are also exploring related information-encoding applications based on these tunable spin–orbit states, which will be reported separately.

 \section{Methods}

\subsection{Experimental}

\begin{figure*} [t]
    \centering
\includegraphics[width=\linewidth]{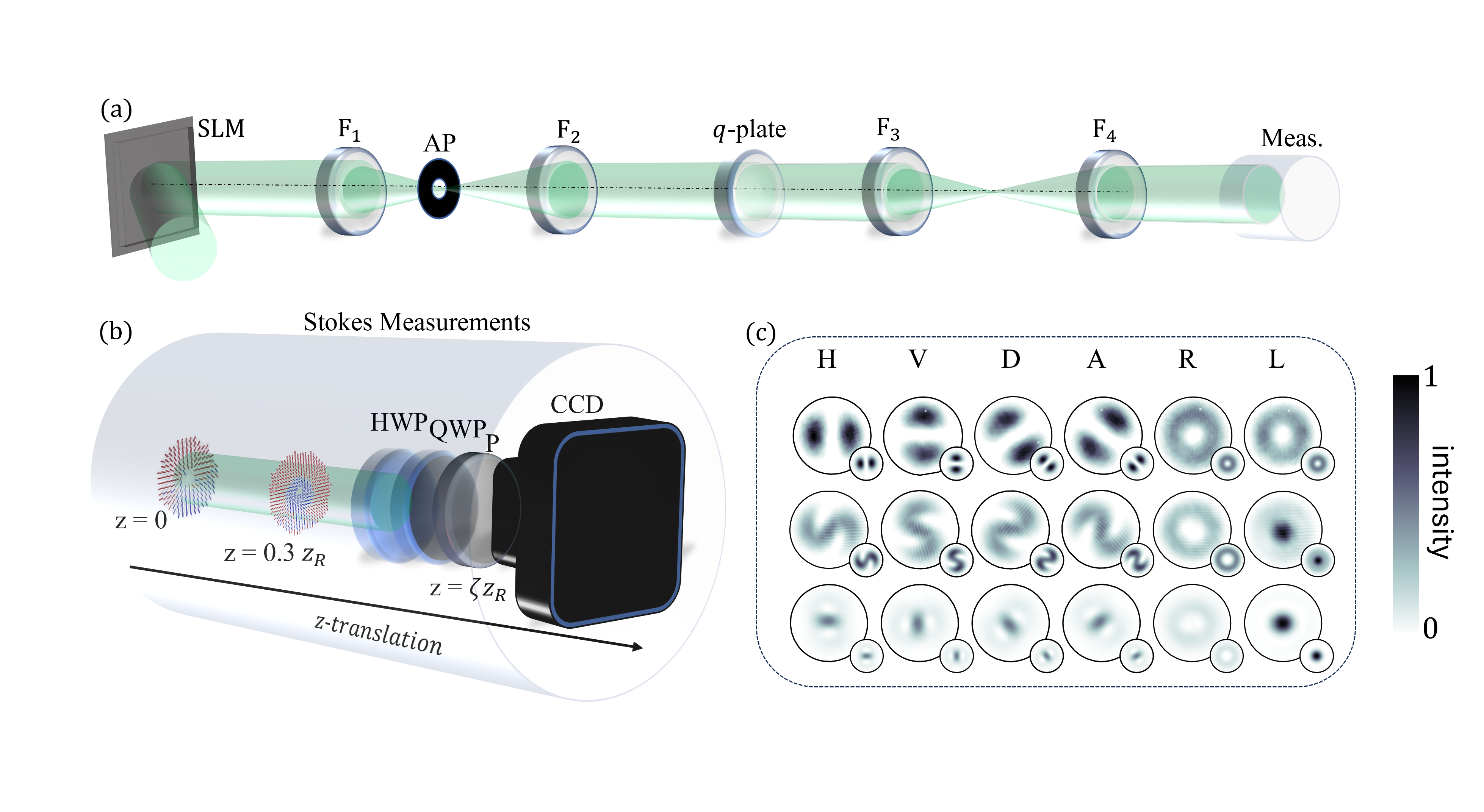}
  \caption{ \textbf{Experimental Setup:}  Schematic of the experimental setup depicting (a) the generation and (b) detection components. SLM: spatial light modulator, AP: aperture, HWP: half-wave plate, QWP: quarter-wave plate, P: polarizer, CCD: charged coupled device. (c) Experimentally recorded polarization intensities for $H, V, D, A, R$ and $L$ for 3 separate propagation planes (inserts denote corresponding theoretical intensity profiles). $\zeta =0, 0.5 \text{ and} $ the farfield. }
    \label{fig:Exp_figure1}
\end{figure*}

We validated our theoretical prediction that the Pancharatnam topological charge gives rise to chirality on the propagation of the optical field through the experimental setup shown in Fig. \ref{fig:Exp_figure1}. The experimental setup consisted of three parts, namely: (1) the generation, (2) propagation and (3) detection (via Stokes polarimetry). In the generation stage, a horizontally polarized Gaussian beam from a HeNe-663 nm laser was expanded and collimated to illuminate a reflective HOLOEYE PLUTO-2.1 LCOS SLM. This scalar Gaussian mode was structured using a LG complex-amplitude hologram with topological charge of $\ell_p \in \{1, -1, 2, -2\}$ encoded on the SLM [top left corner of Fig. \ref{fig:Exp_figure1} (a)], thus producing a LG mode with a phase of $\text{e}^{i\ell_p\phi}$. A 4f-imaging system consisting of lenses F$_1$, F$_2$ = 300 mm was used to image the plane of the SLM onto a $q$-plate.  Since the SLM produces multiple diffraction orders, a spatial filter or aperture (AP) was placed in the focal plane of F$_1$ to select only the positive first order and filter out the remaining orders. The selected first order LG scalar mode was then imaged onto the focal plane of F$_2$ where the $q$-plate was located. The $q$-plate is capable of generating a vector vortex beam by introducing a geometric phase through the variation of the input beam’s polarization, which then imparts OAM. The $q$-plate (in our case having $q = \frac{1}{2}$) converted the incoming horizontally polarized scalar mode into a radially polarized vector mode. Here the horizontally polarized light ($\ket{H}$) can be written as a superposition of  right- and left-handed CP, ($\ket{R}$ and $\ket{L}$, respectively) with the $q$-plate converting $\ket{L}$ into $\ket{R}$ and vice-versa, while introducing a phase of $\text{e}^{i(\ell-2q)\phi}$ and $\text{e}^{i(\ell+2q)\phi}$ to $\ket{R}$ and $\ket{L}$, respectively. A second 4f-imaging system, comprising of lenses F$_3$, F$_4$ = 100 mm, imaged the plane of the $q$-plate to the focal plane of lens F$_4$. Between lens F$_4$ and the charge-coupled device (CCD), we performed Stokes measurements which are further explained below. The CCD was propagated across six different Rayleigh ranges ($z_R$) with the Stokes measurements performed at each individual plane. The components for the Stokes measurements are shown in Fig. \ref{fig:Exp_figure1} (b), with example polarization intensities for 3 separate propagation planes shown in each row of Fig. \ref{fig:Exp_figure1} (c).

\subsection{Stokes parameters and  measurements}

 The components of the locally normalized Stokes vector can then be expressed as 
\begin{equation}
\mathbf{S} \left(\mathbf{r} \right)  = \begin{pmatrix}
    \cos( \Phi) \sin( \beta )\\  
    \sin( \Phi ) \sin(\beta)\\ 
    \cos( \beta )
\end{pmatrix},
\label{eq:skyrmion_sphere}
\end{equation}
where $\Phi$ and $\beta$ are the azimuthal and zenith coordinates on the Poincaré sphere. 

Here, we will give an overview of the representation of the Stokes parameters and the measurement approach used. The reader is advised to view other sources of Stokes polarimetry literature for more detailed and in-depth explanations \cite{singh2020digital}. The four Stokes parameters can be written in terms of the six polarization intensities, namely $I_H, I_V, I_D, I_A, I_R$ and $I_L$, as follows \cite{goldstein2015polarization}:
\begin{align}
    S_0 = I_R + I_L,\\
    S_1 = I_H - I_V,\\
    S_2 = I_D - I_A,\\
    S_3 = I_R - I_L.
\end{align}
$H, V, D, A, R$ and $L$ denote horizontal, vertical, diagonal ($45^\circ$), anti-diagonal ($135^\circ$), right- and left-handed CP, respectively. Three optical elements, namely a polarizer and two retarders (in the form of a half-wave plate (HWP) and  quarter-wave plate (QWP)) can be used to measure the above-mentioned six intensity measurements [as shown in Fig. \ref{fig:Exp_figure1} (b)]. The four linear polarization intensities ($I_H, I_V, I_D$ and $I_A$) can be measured with a polarizer (for $I_H$ and $I_V$) and a HWP and polarizer (for $I_D$ and $I_A$) as the polarizer’s transmission axis transmits the various polarization components in accordance with $I(0^\circ) = I_H$ and $I(90^\circ) = I_V$, while the HWP rotates $I_D$ to $I_H$ and $I_A$ to $I_V$ . By introducing a QWP, which produces a phase shift of $\frac{\pi}{2}$ between the $x$ and $y$ component of the electric field, one can measure circular intensities ($I_R, I_L$), as follows: $I(45^\circ, 90^\circ) = I_R, I(135^\circ, 90^\circ) = I_L$. Therefore, by placing a QWP (set at $90^\circ$) before a polarizer (set initially at $45^\circ$ and then $135^\circ$) the circular intensities ($I_R, I_L$) can be measured.

\section*{Acknowledgments}
The authors thank Andrew Forbes and Ram Khumar for their insightful discussions. I.N. acknowledges support from the South African Quantum Technology Initiative (SAQuTi). I.N. and A.D. gratefully acknowledge funding from the National Research Foundation (NRF) and the Rental Pool Programme (RPP) of South Africa.

\section*{Data availability}
The data that support the findings of this study are available from the
corresponding author upon reasonable request.

\section*{Conflict of interest}
The authors have no conflicts to declare.

\section*{Contributions}
K.A.F. conceived the idea, and I.N. conceived the experiment. I.N and A.D supervised the experiment. L.M. performed the experiment and collected the data, while L.M., together with P.O., carried out the data analysis.  All authors contributed to the writing and revisions of the manuscript.

\bibliographystyle{unsrt}
\bibliography{references.bib}

\end{document}